\documentclass[aps,pre,twocolumn,superscriptaddress,showkeys,showpacs]{revtex4-1}
\usepackage{graphicx}
\usepackage{xcolor}
\usepackage{amsmath}
\usepackage[caption=false]{subfig}
\usepackage{gensymb}

\newcommand{\hl}{\textcolor{black}}

\begin{document}

\title{Influence of media disorder on DNA melting: a Monte Carlo study }

\author{Debjyoti Majumdar}
\email{debjyoti@post.bgu.ac.il}
\affiliation{Alexandre Yersin Department of Solar Energy and Environmental Physics, Jacob Blaustein Institutes for Desert Research,\\
Ben-Gurion University of the Negev, Sede Boqer Campus 84990, Israel}

\date{\today}

% ABSTRACT
\begin{abstract}
We explore the melting of a lattice DNA \hl{in the presence of atmospheric disorder, which mimics the crowded environment inside the cell nucleus, using Monte Carlo simulations. The disorder is modeled by randomly retaining lattice sites with probability $p$ while diluting the rest, rendering them unavailable to the DNA. By varying the disorder over a wide range from $p=1$ (zero disorder) up to the percolation critical point $p_c=0.3116$, we show} the melting temperature $(T_m)$ to increase nearly linearly with disorder \hl{up to $p\approx 0.6$, while strong non-linearity enters for $p\lesssim 0.6$. Associated changes in the bubble statistics have been investigated, showing a substantial change in the bubble size exponents at corresponding melting points for $p\leq 0.5$}. Based on these findings two distinct disorder regimes showing weak and strong effects on melting are identified. For simulations, we use the pruned and enriched Rosenbluth method in conjunction with a depth-first implementation of the Leath algorithm to generate the underlying disorder.
\end{abstract}
\maketitle

% INTRODUCTION
\section{Introduction \label{intro}}

\hl{Free space is limited and energetically costly in the small-scale biological world, leading to molecular crowding. Such an environment compels different functional units to work together in close proximity, in contrast to the isolated conditions often used in \textit{in vitro} experimental setups studying biological processes.} \hl{Crucial examples include biological systems like DNA, which is surrounded by other intracellular} components such as proteins, lipids, saccharides, and other solutes \hl{resulting in a highly crowded environment} \cite{fulton1982, miyoshi2008, skora2020, singh2022, neha2024}. These macromolecular biomolecules occupy about 20-40\% of the cellular volume, which can modify DNA functionality simply by restricting the spatial volume available to the DNA, \hl{thereby playing the role of atmospheric disorder \cite{comm1}}. \hl{Among others, DNA melting is one such example, which is driven by entropic advantage over the energetically favorable base pairing and, therefore, is sensitive to the presence of macromolecular crowders.} 

\hl{Several theoretical \cite{ liu2010, singh2017,hong2020} and experimental studies \cite{woolley1985,nakano2004, harve2009} performed in this direction have shown that the crowders can strongly influence the melting transition. In most cases, a rise in the melting temperature $(T_m)$ with the density of crowders was observed  \cite{woolley1985, harve2009}.} Crowders can also increase the renaturation rate by 1-2 orders of magnitude \cite{wieder1981, sikorav1991}. In some cases, however, a decrease in the $T_m$ was observed too \hl{when low molecular weight  polyethylene glycol (PEG) is used as a crowding agent} \cite{nakano2004}, while larger crowders have been shown to result in a higher increase in $T_m$ \cite{nakano2004, liu2010}. Experimentally, Harve et al. \cite{harve2009} found an enhancement of $7$--$8$\degree C in $T_m$ \hl{while also promoting nucleotide matches}.  \hl{Also studied is the case of a triple-stranded DNA, where the presence of a third strand enhances the stability of the bound state to a greater extent in the presence of PEG, as compared to duplex DNA \cite{goobes2003}.} In a recent theoretical work using the Peyrard-Bishop-Dauxois (PBD) model \cite{pbd1989}, Singh et al. \cite{singh2017} found a linear relationship between the $T_m$ and the crowder density, where the crowder density is varied up to 14\%. 

\hl{In a nutshell, while findings from the studies above show that there is a consensus that $T_m$ increases with the crowder density, it is still unclear how  $T_m$ would behave over a wider range of background disorders, which significantly reduces the free volume available to the DNA. Also, it is unknown if disorder affects the nature of the melting transition, which leaves significant scope for further work in this direction. Taking advantage of this knowledge gap, we aim to study these aspects in this paper using a simple lattice-based model for the DNA and macromolecular crowders.}

For systems defined on the lattice, the usual way to realize a disordered background is to use percolation-type models where sites diluted with a certain probability are rendered non-functional or differently functional than their typical behavior in the non-diluted system. Whereas a large number of studies in the last few decades has been devoted towards understanding the changes in the polymer scaling laws in disordered lattices \cite{bublee1988,meir1989,rintoul1994,singh2009,blavatska2010}, studies concerning disorder-induced changes in DNA melting using lattice-based models, remains less explored. Of particular importance is the question of whether a melting transition exists at all in the limit where the fraction of available sites is close to the percolation threshold and the underlying lattice is a fractal characterized by broken dimensions \cite{stauffer1992}.  The additional diverging length scale at the percolation critical point is expected to make things complicated, which demands further attention and careful study using simplistic but versatile DNA models, which can be easily integrated with models of percolation and allow usage of powerful numerical techniques at the same time. 

Additionally, melting on the fractal infinite cluster at the percolation threshold has some special relevance per se since the chromatin in its compact form exhibits fractal-like properties, with a fractal dimension $d_f=2.4$  as revealed from small angle neutron scattering experiments \cite{metze2013}. This fractal form gives rise to anomalous properties, e.g., sub-diffusive dynamics of chromosomal loci \cite{tamm2015}, not only with active forces in a non-equilibrium backdrop but also for the thermal equilibrium scenario \cite{weber2011, singh2024,majumdar2024}. \hl{However, it is essential to mention that the fractal chromatin arises due to non-equilibrium effects; therefore,} the scenarios concerning the fractal form of chromatin and melting on a fractal lattice \hl{are not directly related}. However, it is still plausible that the underlying fractal structure preserves some universal features that would be reflected in both situations.

In this paper, we present results for simulations of DNA melting on the infinite cluster backbone at the site percolation threshold $(p_c=0.3116)$ of the three-dimensional cubic lattice and also for other values of disorder $(p\geq p_c)$, using a lattice adaptation of the Poland-Scheraga (PS) model \cite{causo2000, poland1966} of the DNA. The phase diagram demonstrating how the melting temperature varies with the degree of disorder is mapped out, and the changes in the associated scaling exponents and order parameter distribution at the transition points are investigated.  Further, we also study the bubble formation statistics, which is believed to be related to crucial functionalities of the DNA, from providing flexible hinges to fold \cite{geggier2010,yuan2006} to initiation of transcription \cite{titus2005}. Below $p_c$, the clusters are disconnected such that they cannot support a chain of infinite length (thermodynamic limit), and, therefore, the question of a phase transition is moot. Other than the melting transition, we report possible enhancements in the numerical algorithm, which could enhance the sampling of polymers in disordered media. 

The rest of the paper is organized in the following manner: in Sec. \ref{model}, we introduce the models for DNA and lattice disorder. Sec. \ref{simtech} discusses the simulation techniques for introducing lattice heterogeneity and growing the DNA strands on it. In Sec. \ref{scaling}, the observables of interest, the associated scaling forms, and the method of disorder averaging are discussed. In Sec. \ref{results}, we discuss the findings on how a disordered environment modifies the DNA melting transition with particular emphasis on the bubble statistics, and finally conclude our paper in Sec. \ref{conclusion}. 

% MODEL 
\section{Our Model for DNA and disorder\label{model}}

\begin{figure}[t]
\includegraphics[width=.9\linewidth]{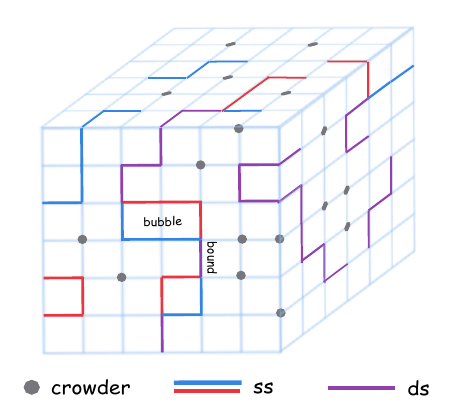}
\caption{Schematic depiction of our model DNA on a three-dimensional slice of the cubic lattice with crowders. ss (ds) denotes the single (double) stranded segments. Terminating ends of the strands denote moving to the next plane.}
\label{dnafig}
\end{figure}

\textit{DNA model:} We \hl{consider a} lattice model of a homogeneous DNA \hl{in the dilute limit where only a single DNA molecule is present}. Two distinct self-avoiding walks $(\textbf{r}^A$ and $\textbf{r}^B~)$ originating from the center of a cubic lattice of linear dimension $L$, represent the double strands of the DNA [Fig. \ref{dnafig}]. Besides being self-avoiding, the strands are also mutually-avoiding. The only exception is for monomers with the same position index along the strands, which can occupy the same lattice site $(\textbf{r}_i^A=\textbf{r}_i^B)$ resulting in an energetic gain of $-\epsilon$, thereby mimicking the hydrogen-base pairing in DNA. \hl{One end of the DNA is pinned at the origin, while the other end is free to wander. Essentially, this model comprises the following key features: double-stranded bound segments, unbound segments called bubbles, and a Y-fork at one end.} The Hamiltonian describing a typical configuration would be $\mathcal{H} = -\epsilon\sum_{i=1}^N \delta_{\textbf{r}_i^A, \textbf{r}_i^B}$, where $\delta_{i,j}$ is the Kronecker delta counting the number of base-pair contacts \hl{and $N$ is the maximum number of possible base pairs}. With every base pairing, we associate a Boltzmann factor $\exp(\epsilon/k_BT)$, where $T$ is the temperature and $k_B$ is the Boltzmann constant. We set $\epsilon=k_B=1$ throughout our simulations. From here onwards, we will refer to $N$ as the DNA's length or system size. \hl{Our model is a lattice adaptation of the famous PS model \cite{poland1966} and was introduced in Ref. \cite{causo2000} and later used for studying multiple scenarios of DNA melting \cite{majumdar2020,majumdar2021,majumdar2023,majumdar2023p2}, including the effect of sequence heterogeneity \cite{coluzzi2006}}.

\textit{Crowder model:} We randomly dilute sites across the cubic lattice, thereby introducing  background disorder to model the macromolecular crowders [Fig. \ref{dnafig}]. Diluted sites are then no longer available to grow the DNA chain/s. The defects generated this way are, therefore, spatially uncorrelated and, the fraction of diluted sites correspond to the crowder density. However, modeling this way, we assume crowders of uniform sizes only, which are frozen in time or with a relaxation time much larger than the time the DNA  would take to sample different regions of the available volume within the observation time. 

Our model, therefore, simplifies the actual complex situation by coarse-graining microscopic details at different levels, e.g., we neglect the sequence heterogeneity along the DNA strands, the difference in the bending rigidity among the bound and unbound segments, helical topology, polydispersity of the crowders, etc.  While our model cannot explicitly include some of these features, e.g., helicity, including others, would only make the problem infeasible to study along with a disordered background. Therefore, we plan to consider some of these separately in future works.   

%SIMULATION TECHNIQUE
\section{Simulation techniques \label{simtech}}

% LATTICE DISORDER GENERATION
\subsection{Lattice disorder generation method}

We use the Leath algorithm \cite{leath1976} to generate the infinite cluster at the site percolation threshold $(p_c)$ and also for other disorder values $p>p_c$. Starting from the center of a cubic box, which we assume to be occupied and later serve as the starting point of the DNA configurations, the neighboring sites are visited and occupied with probability $p$.  \hl{Since there are no special directions, we can search the neighbors in any sequence.} If chosen unoccupied, we still mark the site as visited so that it will not be considered for occupancy in the future to ensure random but uniform dilution of sites. If the recent site is occupied, we further perform a recursive depth-first search and occupy for the neighbors of the current site. This process continues as long as the pointer does not try to step out of the simulation box or there are no more available unvisited neighboring sites.  Once stuck, the pointer returns to the last occupied site and continues with its other neighbors. To ensure that an infinite cluster exists ($p=p_c$), we check if the cluster being generated touches all six faces of the cube. If the cluster fails to connect all of the 2D faces \hl{(D is the dimensionality),} we discard the current realization and start with a new one. \hl{Further, we have also checked with the breadth-first approach of generating the underlying disorder. The breadth-first implementation offers better statistics, at least for lattice animals \cite{hsu2005}, due to the lower fluctuation in growth sites. It is, therefore, important to check if the same holds for the DNA melting problem.}

To avoid boundary effects of the finite simulation box containing the disordered lattice, we used lattices of linear dimension $L=599$, much larger than that would be required by SAWs even at $p_c$ with a modified size exponent $\nu_{\text{SAW}}(p_c)=0.667$ \cite{blavatska2010}. \hl{We use bit-map to encode the lattice occupation using the following rule: an unavailable disordered site is indexed as `$0$', an available but unoccupied site is indexed as `$1$', and an available but occupied site is indexed as `$2$'.} To ensure the fractality of the infinite cluster at the percolation threshold, we calculate the mass fractal dimension $(d_f)$ given by the scaling of the number of sites with the radius $(r)$ of concentric circles, $M\sim r^{d_f}$, with the center chosen at the cluster's center of mass. For three-dimensional site percolation, the value is precisely known to be $d_f=2.52$ \cite{stauffer1992}, which matches well with our estimate.

% POLYMER GENERATION
\subsection{Polymer generation method}

To simulate the DNA strands on the diluted lattice obtained using the method mentioned in Sec. \ref{simtech}(A), we use the pruned and enriched Rosenbluth method (PERM) \cite{grassberger1997, bachmann2004} \hl{which presents a considerable improvement on the Rosenbluth-Rosenbluth (RR) method \cite{rosenbluth1955}. PERM employs the RR method along with population control, significantly increasing the number of successfully generated chains at long lengths. While PERM was originally introduced for a single chain \cite{grassberger1997}, extending to multi-chain systems, like DNA \cite{causo2000}, is straightforward, as is discussed below.}   

\hl{Starting from the origin of a cubic lattice,} two strands of the DNA are grown simultaneously, while monomers are added to the \hl{growing end} of both the strands at once. At each step, we calculate the combined possibilities of free sites to step \hl{into}, obtained by a Cartesian product of the individual sets of free sites for each strand, i.e., $\mathcal{S}_n=\mathcal{S}_n^A \times \mathcal{S}_n^B$, where $\mathcal{S}_n^A$ and $\mathcal{S}_n^B$ are the individual sets of possible free sites. Each element in $\mathcal{S}_n$ represents an ordered pair of new steps for both the strands, and the importance is given by the Boltzmann weight $\exp(\epsilon\beta)$ for a base-pair contact and $1$ otherwise. A choice is made by picking a uniform random number $\in[0,w_n]$, where $w_n=\sum_{\mathcal{S}} \exp(\beta \epsilon \delta_{r_n^A,r_n^B})$ is the one-step local partition sum \hl{for $n$th step}, and then finding the $\mathcal{S}_n$ element it corresponds to. The current weight at length $n$ is given by the product of the local partition sums at each step, $W_n=\prod_{i=1}^n w_i$. Averaging $W_n$ over the number of started tours then gives the average partition sum, $Z_n=\langle W_n \rangle$, \hl{where a `tour' is a collection of chains created between two successive returns to the main subroutine}.

Population control at each step is performed by enriching with configurations of higher weights and pruning configurations of smaller weights probabilistically. \hl{This is achieved by recursive calls to the PERM subroutine} depending on the ratio, $r=W_n/Z_n$:

\begin{equation}
r=\begin{cases}
<0.9, & \text{prune with probability $(1-r)$} \\
[0.9,1.1], & \text{continue to grow} \\
>1.1, & \text{make $k$-copies.}
\end{cases}
\label{rateq}
\end{equation}

If $r<0.9$ and pruning fails, the configuration is continued to grow but with $W_n=Z_n$. For enrichment ($r>1.1$) $k$ is chosen as, $k=min(\lfloor r \rfloor,\mathcal{N}(S_n))$, where $\mathcal{N}(\mathcal{S}_n)$ is the cardinality of the set $\mathcal{S}_n$, and each copy is continued with a reduced weight $\frac{W_n}{k}$ . \hl{A tour, therefore, has a rooted tree topology where the growth along a single branch is continued up to the maximum length $N$ or until it is pruned, while the tour's growth continues as long as branches are left to grow.}

\hl{For generating uniform random numbers, we used the Mersenne-Twister (MT)  random number generator (RNG) as implemented by Matsumoto and Nishimura \cite{matsumoto1998}. We also checked with other RNGs like RAN2 from numerical recipes  \cite{numericalrecipes} and found MT to be at least three times faster than the RAN2. }

% ADDITIONAL BIAS
\subsection{Additional bias}

One of the ways to avoid polymer growth from getting stuck in constrained geometries, e.g., on cylinder, is to use \textit{Markovian anticipation} \cite{frauenkron1999} where depending upon $k$--steps statistics at length $m$, we decide what should be the choice for the future step at length $n$ depending upon the sequence of $(n-1-k)\cdots (n-1)$ steps. However, we must prepare the initial set of $k$--step statistics for each disorder realization to apply such a bias for growth in a disordered lattice. Hence, we need a scheme that can bias only depending on the local density of diluted and occupied sites. Thus, in addition to associating weights to base-pair contacts, to favor the growth of chains towards a less diluted \hl{and empty} zone, we apply an extra directional bias in which the next steps of the walkers are biased in the direction of the pyramidal cone formed by the $h$ successive layers with the growing end forming the apex, corresponding to each $s_n\in \mathcal{S}_n$, which has the lowest diluted and occupied sites. The weight used for such a bias is of the form $f_{bias}=\frac{n_{as}}{(n_{os}+1)}$ , where $n_{as}$ and $n_{os}$ are the number of total available sites and the number of occupied sites within the volume of the pyramid, respectively, for each $s_n\in \mathcal{S}_n$. Note that the $1$ in the denominator avoids divergence if $n_{os}=0$.  The depth of the pyramid determines how far the walker sees before taking the next step. The dimensions of the scanning pyramid are determined by the height $h$, and the base, which is a square of width $2h$. The time required to scan the pyramidal volume increases like $\mathcal{O}(h^3)$. In our simulations, we use $h=3$. With the introduction of this extra bias, the expression for calculation of weights at each step has to be modified as $w_n=\sum_{j\in\mathcal{S}} \exp(\beta \epsilon \delta_{r_n^A,r_n^B})f_{bias}^j$. Of course, at each biased step, the inclusion of the local weights needs to be corrected by the extra biasing factor corresponding to the direction of the chosen steps, i.e., $W_n/f_{bias}^j$, where $f_{bias}^j$ is the biasing factor for the chosen pair of directions. \hl{Using this additional bias, we observed a two-fold increase in the number of walks reaching length $N$ at long times.}

% SCALING AND AVERAGING
\section{Observables, scaling and averaging \label{scaling}}
\hl{Estimate of thermodynamic averages begin with estimating the partition function which contains the weighted sum over all possible states, $Z_n=\sum_i g(E_i) e^{-\beta E_i}$, where $g(E_i)$ denotes the density of states with energy $E_i$. Thereafter, the expectation value of any observable (say $ Q_n$) at length $n$, is simply given by}
\begin{equation}
\langle Q_n \rangle (T) = \frac{\langle Q_n W_n(T) \rangle}{Z_n(T)},
\label{qnavg}
\end{equation}
where the $\langle \cdots \rangle$ in the numerator represents the running average of the quantity over the number of started tours, using the local estimates of the configuration weight $W_n(T)$. \hl{Besides}, we also need to perform disorder averaging, which we will discuss in the upcoming paragraphs.

To study the DNA strand separation transition, we look at the average number of bound base-pairs $(\langle n_{c} \rangle)$ at different temperatures, which also serves as the order parameter and the average energy. Under a change of temperature from \hl{$T=0$ to $T=\infty$}, $n_{c}$ goes from $n_{c}/N=1$ (bound) to $0$ (unbound) phase.  Around the transition point, we have the following scaling 
\begin{equation}
n_c(T)=N^\phi h[(T-T_m)N^\phi]
\label{nceq}
\end{equation}
\hl{where the exponent  $\phi$ controls the sharpness of the transition, and $h(x)$ is some scaling function}. For first-order melting transitions $\phi=1$, and $\phi=1/2$ for continuous melting transitions (e.g., for ideal chains) \cite{causo2000}. The thermal response is obtained from the fluctuation of $n_c$ and is given by $C_c=\langle n_c^2 \rangle - \langle n_c \rangle^2$, where the $\langle \cdots \rangle$ denotes averaging over configurations. The quantity $C_c$ follows the scaling form 
\begin{equation}
C_c(T)=N^{2\phi}g[(T-T_m)N^\phi]
\label{cceq}
\end{equation}
 near the transition point. Therefore, we will get data collapse on plotting $n_c/N^{\phi}$ or $C_c/N^{2\phi}$ vs. $(T-T_m)N^\phi$ using which one can extract the melting points $T_m$ and exponent $\phi$. 

\hl{To verify changes in nature of the melting transition}, we find the bubble size distribution $P(\ell_b)$ at the transition points, where \hl{a bubble is defined to be a contiguous set of broken bonds enclosed within bound segments, and the difference between the bound base-pairs indices enclosing the bubble corresponds to $\ell_b$.} The bubble size has been shown to follow a power law distribution of the form $P(\ell_b)\sim \ell_b^{-c}$, where $c$ is called the bubble size or the reunion exponent. For first-order transition $c\geq 2$, and $1<c<2$ for continuous transition. Further, for continuous transitions, we have $\phi=c-1$ \cite{carlon2002}. Note that, due to the lattice's discrete nature, the minimum size of a bubble starts from $\ell_{b,min}=2$. \hl{Other than $\ell_b$, we also studied the average number of bubbles $(n_b)$ below the melting transition}.

We also looked at the base-pair contact probability distribution $(P_{n,n_c}(T))$ at different lengths close to the transition points. To calculate $P_{n,n_c}(T)$ we use the following formula
\begin{equation}
P_{n,n_c}(T) = \frac{Z_{n,n_c}(T)}{\sum_{n_c=0}^n Z_{n,n_c}(T)},
\label{pnceq}
\end{equation}
where $Z_{n,n_c}(T)$ is the constrained partition sum at length $n$ with $n_c$ number of base-pair contacts. For the DNA model in hand, the probability distribution is expected to follow the scaling form $P_{n_c,N}\sim N^{-\phi}p(n_c/N^\phi)$ \cite{causo2000}. 

Besides the averaging performed over multiple tours (thermal fluctuations) for a given instance of disorder -- denoted by $\langle \cdots \rangle$ -- we also need to perform averaging over distinct disorder realizations, which we denote by $\left[\cdots\right]$. Doing so, we notice that the disorder averaging of an observable $\left[ \langle Q_n\rangle \right]$ can be done at two different levels; first, the average can be taken over $\left[ Z_n \right]$ which would give us the  annealed free energy $f_a=-\beta ^{-1}\ln  \left[ Z_N \right]$ and the expression for evaluating a disorder averaged observable will then be given by,

\begin{equation}
 \left[ \langle Q_n\rangle \right]_a=\frac{\left[ \langle Q_n W_n \rangle \right]}{\left[ Z_n  \right]}.
 \label{anavg}
\end{equation}

In the second kind of average, the disorder averaging is taken over the logarithm of $Z_n$ as $\left[ \ln Z_n \right]$,  which gives the quenched free energy $f_q=-\beta ^{-1} \left[ \ln Z_N \right]$. Here, an averaged observable $ \left[ \langle Q_n\rangle \right]$ will be given by,

\begin{equation}
\left[ \langle Q_n \rangle\right]_q= \left[  \frac{\langle Q_n W_n \rangle}{ Z_n} \right]=\frac{1}{C} \sum_{C} \frac{\langle Q_n W_n \rangle}{Z_n},
\label{qavg}
\end{equation}
where $C$ is the number of independent disorder configurations generated, with at least one DNA sample of length $n$ for the $\langle \cdots \rangle$ average. The way PERM is implemented, it is perhaps easier to implement Eq. \ref{anavg}. Again, since one of the ends of our model DNA remains pinned at the origin, one can argue that our study corresponds to the ``quenched" problem \cite{doussal1991}. 

While averaging over different disorder realizations, the convergence of results can be sensitive to the number of independent disorder realizations $(\eta_1)$, and the number of independent samples (here called ``tours") ($\eta_2$) used for averaging over each disorder realization. For our purpose, we found $\eta_2=10^4$, and \hl{$\eta_1=10^{8}$} to give sensible results. 

\begin{figure}[t]
\centering
\includegraphics[width=\linewidth]{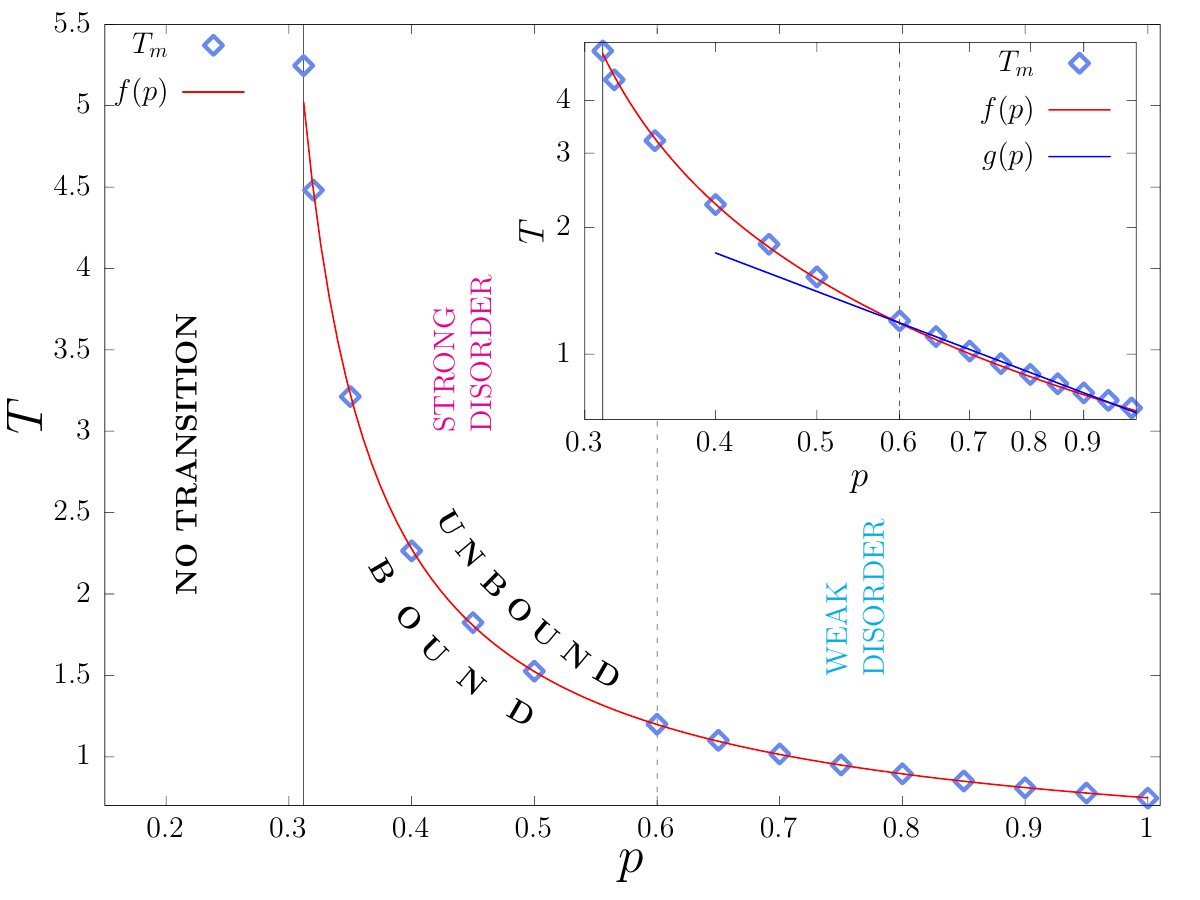}
\caption{\hl{Melting vs. disorder phase diagram showing the variation of $T_m$ with $p$. The dashed vertical line separates the whole range of disorder from $p=p_c$ to $p=1$ into `weak' and `strong' disorder regimes. The regime `no transition' refers to $p<p_c$. $f(p)=1/\ln(c_1 p+c_2)$ is a fit using all the data points, where $c_1$ and $c_2$ are constants.  (Inset) It is the same as the main plot but on a log-log scale. $g(p)\sim p^{-\alpha}$ is a fit to the data points in the range $p\in[0.6,1]$, yielding $\alpha=0.94\pm 0.02$.}} 
\label{pdfig}
\end{figure}

%RESULTS AND DISCUSSION
\section{Results and discussion \label{results}}

%REGULAR LATTICE
\textit{Regular lattice:} The unbinding phase transition of a duplex DNA is the result of an underlying mechanism trying to minimize the free energy density by increasing the entropy of the system according to the equation $ F = U - T S$, where $U$ and $S$ is the energy density and entropy per unit length, respectively, at a constant temperature $T$.  For the model considered here, the melting point on a regular lattice, i.e., $p=1$, is $T_m=0.7454$ \cite{causo2000}, and the melting transition was discontinuous (first-order), with an exponent value $\phi \approx 1$ \cite{causo2000}. Across a first-order melting point $F_{T\rightarrow T_{m^-}}=F_{T\rightarrow T_{m^+}}$, therefore, change in F, $\triangle F\vert_{T=T_m}=0$ yielding $T_m=\frac{\triangle U}{\triangle S}$. One can, therefore, identify the bound phase as the energy-dominated state and the unbound phase as the entropy-dominated state, with $T_m$ determined by an interplay between $\triangle U$ and $\triangle S$. Note that while $\triangle U$ is fixed while going from bound to unbound phase, $\triangle S$ usually depends on the connectivity of the underlying lattice, which will play an important role in the present study.

%PHASE DIAGRAM
\textit{Phase diagram:} \hl{We show the melting phase diagram as a function of lattice disorder in Fig.~\ref{pdfig} and in the log-log scale in Fig.~\ref{pdfig}(inset). The melting temperature $(T_m)$ increases non-linearly with an increase in disorder or decrease in $p$. To fit the data points, we use a fitting function of the form $f(p)=1/\ln(c_1p+c_2)$ where $c_1=3.85\pm 0.05$ and $c_2=0.009\pm 0.01$ are fitting parameters [Fig.~\ref{pdfig}]. A perfect linear variation of $\exp(1/T_m)$ with $p$ motivates this choice of $f(p)$. Interestingly, on fitting the datapoints for $p\in [0.6,1]$ with a function of the form $g(p)\sim p^{-\alpha}$, we get a nearly linear fit with an exponent $\alpha=0.94\pm0.02$. This implies, we can assume an almost linear dependence on $p$ for low disorder values, which is in line with the findings of Singh et al. \cite{singh2017}. For higher disorder pertaining to $p<0.6$, however, the dependence on $p$ becomes strongly non-linear. On top of that, we divide the phase diagram into three regimes; $0.6 \leq p \leq1$ denotes the weak disorder regime, $p_c \leq p \leq0.6$ is the strong disorder regime, and $p<p_c$ as the region for no phase transition since clusters of the underlying lattice are disconnected and an infinite cluster does not exist. The basis of this distinction will become clear as we look into different quantities. }

\begin{figure}[t]
\centering
\includegraphics[width=\linewidth]{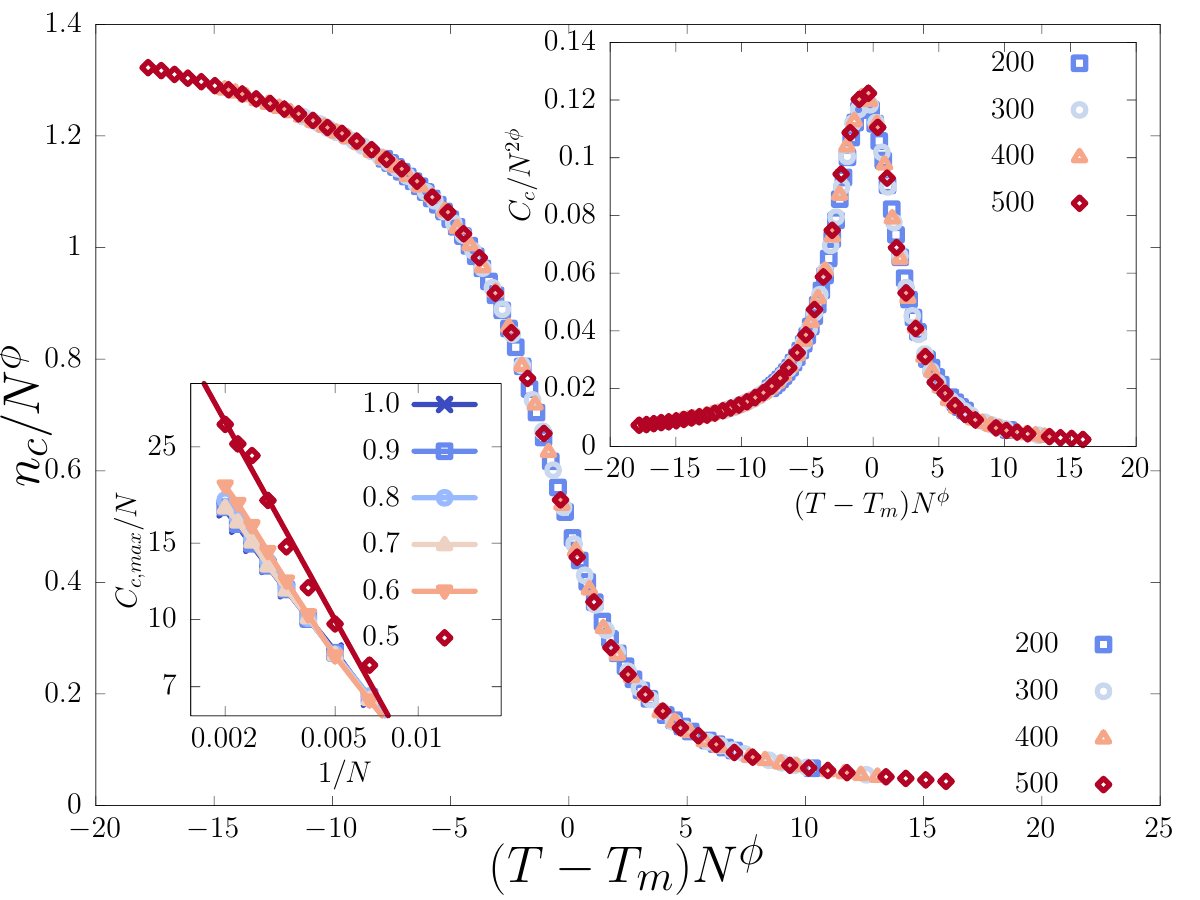}
\caption{Scaling plot of average number of base-pairs in contact for $p=0.8$ using $\phi=0.9$ and $T_m=0.895$. (Right inset) Data-collapse of $C_c$ data for $p=0.8$ using the same $T_m$ and $\phi$ as in the main plot. (Left inset) $C_c$ peak values for  $p=0.5$ to $1$, vs. system size inverse $(N^{-1})$.} 
\label{ncfig}
\end{figure}

The melting points are estimated from data collapse of the order parameter $(n_c)$ and its fluctuation $(C_c)$ [Fig. \ref{ncfig} and inset] at different lengths across the melting transition, extracting the exponent $\phi$ at the same time. However, for values of $p\lesssim0.4$, with growing sample-to-sample fluctuation, the convergence of \hl{the order parameter cumulants }(such as $C_c$), is difficult, which led us to estimate $T_m$ relying only upon the first moment, i.e., $n_c$ itself, \hl{and that too with lesser accuracy}.  We found Eq. \ref{anavg} and Eq. \ref{qavg} to give nearly identical values for $n_c$, with Eq. \ref{qavg} giving slightly smaller $T_m$ for $p\neq 1$, but the same for $p=1$. Similar equivalence \hl{between annealed and quenched averaging} for single polymers in disordered media was pointed out in the past \cite{cherayil1990,wu1991,blavatska2013}, and more recently for semi-stiff polymers in heterogeneous lattices \cite{bradly2021}.

  The increase in $T_m$ occurs due to the lowering of entropy $(S)$ \hl{in the unbound phase} when sites are increasingly less available for lower $p$ values, while $\mid \triangle U\mid=\epsilon$ remains fixed across the melting transition. \hl{The divergence in $C_c$ becomes stronger for $p\lesssim 0.5$; while for $p>0.5$ the way $C_c$ diverges with system size at $T_m$ remains same [Fig. \ref{ncfig}(inset)]. This indicates a stronger effect on the melting transition besides simply changing the $T_m$ and also connects with how the phase line in Fig. \ref{pdfig} shows stronger non-linearity for $p\lesssim 0.6$.} 

\hl{To check that the results are independent of the way the underlying disorder is created, we performed additional simulations using the breadth-first approach to create the infinite cluster. We found both approaches (depth and breadth-first) to give identical results, even though the shape of the produced infinite cluster using these two methods can be drastically different.}   

\begin{figure}[t]
\includegraphics[width=\linewidth]{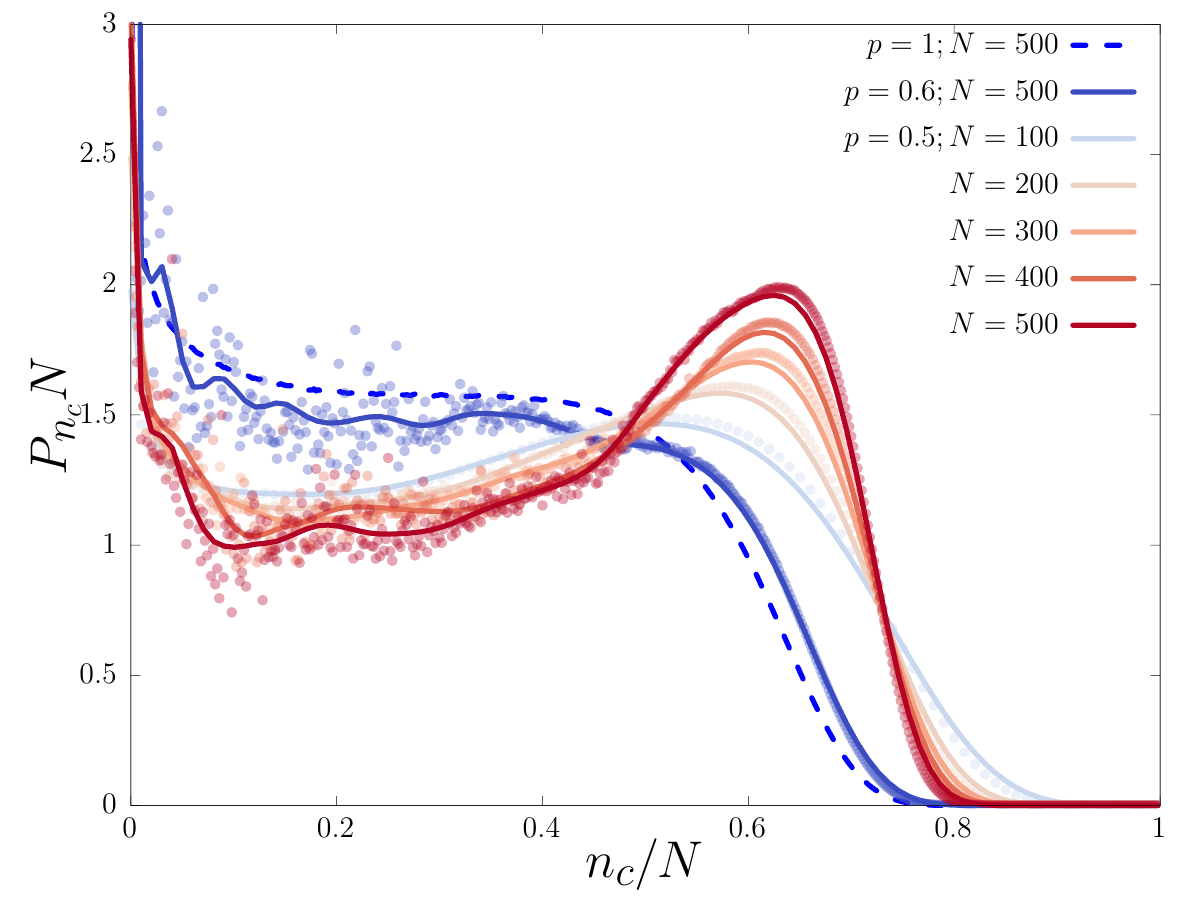}
\caption{Scaling of order parameter distribution $(P_{n_c})$ with system size at the melting point of $p=0.5, 0.6$ and $1$. Data shown for $p=0.5$ corresponds to lengths $N=100$ -- $500$, and $N=500$ for $p=0.6$ (blue solid) and $1$ (blue dashed). For comparison, we have also shown data for $p=1$ at $T_m=0.7454$ for $N=500$ as a dashed line. Solid lines are an approximation of the data with the B\'ezier curve.}
\label{pncfig}
\end{figure}

%ORDER PARAMETER DISTRIBUTION
\textit{Order parameter distribution:} The order parameter per se is not enough to reveal all the crucial details of the melting transition and, therefore, we need to look at its distribution $P_{n_c}(T)$ as well, especially, close to the transition point. In Fig. \ref{pncfig}, we plot $P_{n_c}$ at the melting point for $p=0.5$, comparing data for chain lengths $N=100$ -- $500$, along with the data for $p=1$ and $0.6$ for $N=500$. Here, we take an extra averaging over 20 independent runs. Clearly, even at the system sizes considered, one can easily discern the growing peak at $n_c/N\sim 0.7$, and a deepening valley at  $n_c/N\approx 0.2$ for $p=0.5$, which is absent for $p=1$ or even for $p=0.6$ \cite{causo2000}. The absence of a pronounced peak for $N=100$ denotes \hl{a finite size effect and} that the effect of lattice heterogeneity has not been felt by the DNA strands yet to the extent that it modifies the distribution, and the effect becomes more prominent for longer chain lengths. \hl{Here, too, we see the effect of disorder becoming stronger for $p=0.5$, while $p=0.6$ shows behavior similar to $p=1$ with no additional peak.}

To understand the \hl{change in $P_{n_c}$}, note that a first-order transition is generally characterized by a doubly peaked distribution separated by a valley whose depth grows with the system size $(L)$ as $\exp(-\sigma L^{d-1})$. The valley results from the $d-1$ dimensional surface separating the coexisting phases in the $d$-dimensional Euclidean space embedding the system. For our model DNA, which is topologically one-dimensional, any valley separating the bound and unbound states is absent for  $p=1$ since, while going from a bound segment to an unbound segment along the DNA chain, there is no surface energy-like cost that is extensive and, therefore, the states in-between are not suppressed. On the other hand, for DNA in a sufficiently disordered environment, the emerging valley results from the ensuing entropy crisis, suppressing the intermediate states thereof. For $n_c/N\rightarrow 0$, configurations of two individual strands of effective length $2N$ need to be sampled, resulting in an increased fluctuation of the $P_{n_c}$ curve as compared to the $n_c/N>0.6$ side.

\begin{figure}[t]
\includegraphics[width=\linewidth]{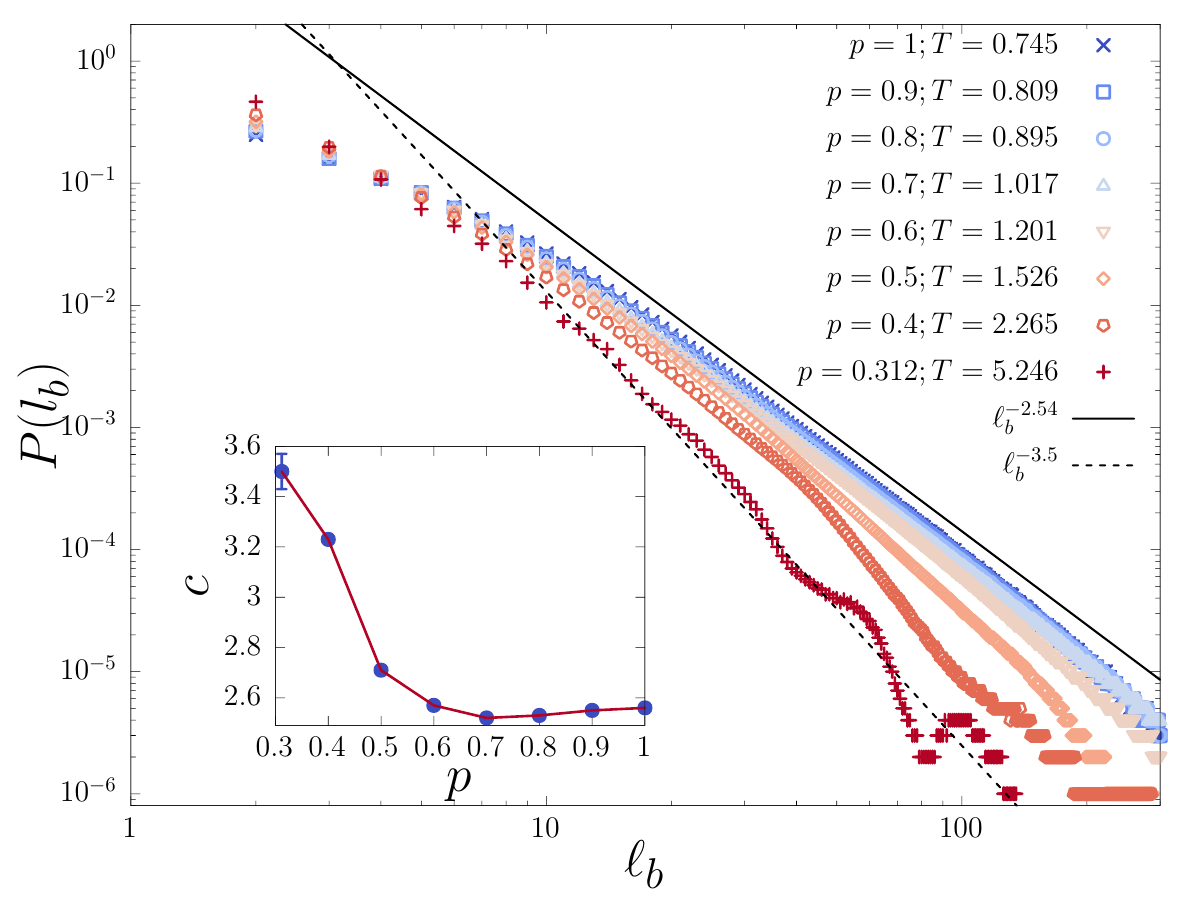}
\caption{Bubble size distribution at the corresponding melting point of systems with disorder values $p=1$ -- $0.312$. The solid line is a power law $P(\ell_b)\sim \ell_b^{-c}$ fit of the $p=1$ data points in the range $\ell_b=20-100$ giving $c=2.54$ \cite{majumdar2020}, and the dashed line for the data points of $p=0.312$ in the same range giving $c=3.5$. (Inset) The variation of the exponent $c$ for different values of $p$.}
\label{figbsd}
\end{figure}

%BUBBLE STATISTICS 
\textit{Reunion or bubble statistics:} Next, we come to bubble statistics, where we find the bubble-size-distribution ($P(\ell_b)$) at the corresponding transition points and the average number of bubbles $(n_b)$ across the transition for different $p$ values. At the melting point, $P(\ell_b)$ follows a power law of the form $P(\ell_b)\sim\ell_b^{-c}$, where $c$ is the bubble size exponent \cite{carlon2002}.  The advantage of measuring $c$ is that it is robust with system size and, therefore, less affected by finite size corrections. 

In Fig. \ref{figbsd}, we plot the BSD  at the corresponding melting points for \hl{ different $p$ values}. The exponent $c$ is extracted by fitting the intermediate data points in the range $\ell_b=20$ -- $100$, which comes out to be $c=2.54 \pm 0.005$ \cite{majumdar2020} for $p=1$ and $c=3.5 \pm 0.06$ for $p=0.312$. \hl{An increase in $c$ with decreasing $p$ is according to our expectation since higher disorder should make the reunion of two strands and, therefore, the formation of larger bubbles difficult}. A similar increase in the exponent for loop formation was found for polymer in correlated disorder in Ref. \cite{haydukivska2016}. In Fig. \ref{figbsd} inset, we show the variation of the exponent $c$ with $p$. Noticeably, the major observable change in $c$ occurs mainly for $p\lesssim 0.6$, while it remains almost the same above $p=0.6$. \hl{This trend of change in $c$ with $p$ is akin} to what we found \hl{for the scaling of $T_m$ with $p$ [Fig. \ref{pdfig}], for the scaling of $C_c$ peaks [Fig. \ref{ncfig}(inset)], and for $P_{n_c}$ [Fig. \ref{pncfig}]}.  On this basis, we \hl{demarcate} the effect of disorder on DNA melting into two distinct regimes: the `weak' disorder regime for $p \in [0.6,1]$, and the `strong' disorder regime for $p \in [0.312,0.6)$, \hl{where $p=0.6$ is only a rough estimate of the separating point and require further simulations for accurate determination.} 

The undulant behavior at larger bubble sizes for $p=0.4$ and $0.312$,  which is absent for higher $p$ values, is because, as $p$ approaches $p_c$, entropically rich regions within the infinite cluster are connected by entropically unfavorable regions (bottlenecks) which forces the two strands to reunite \hl{while passing between two strongly connected clusters} giving rise to intermittent rise in $P(\ell_b)$ value. \hl{This further suggests that not all parts of the DNA experience the same environment (entropy), and melting happens heterogeneously along the chain with the bulk melting temperature given by an averaged value over the chain.}

\begin{figure}[t]
\includegraphics[width=.9\linewidth]{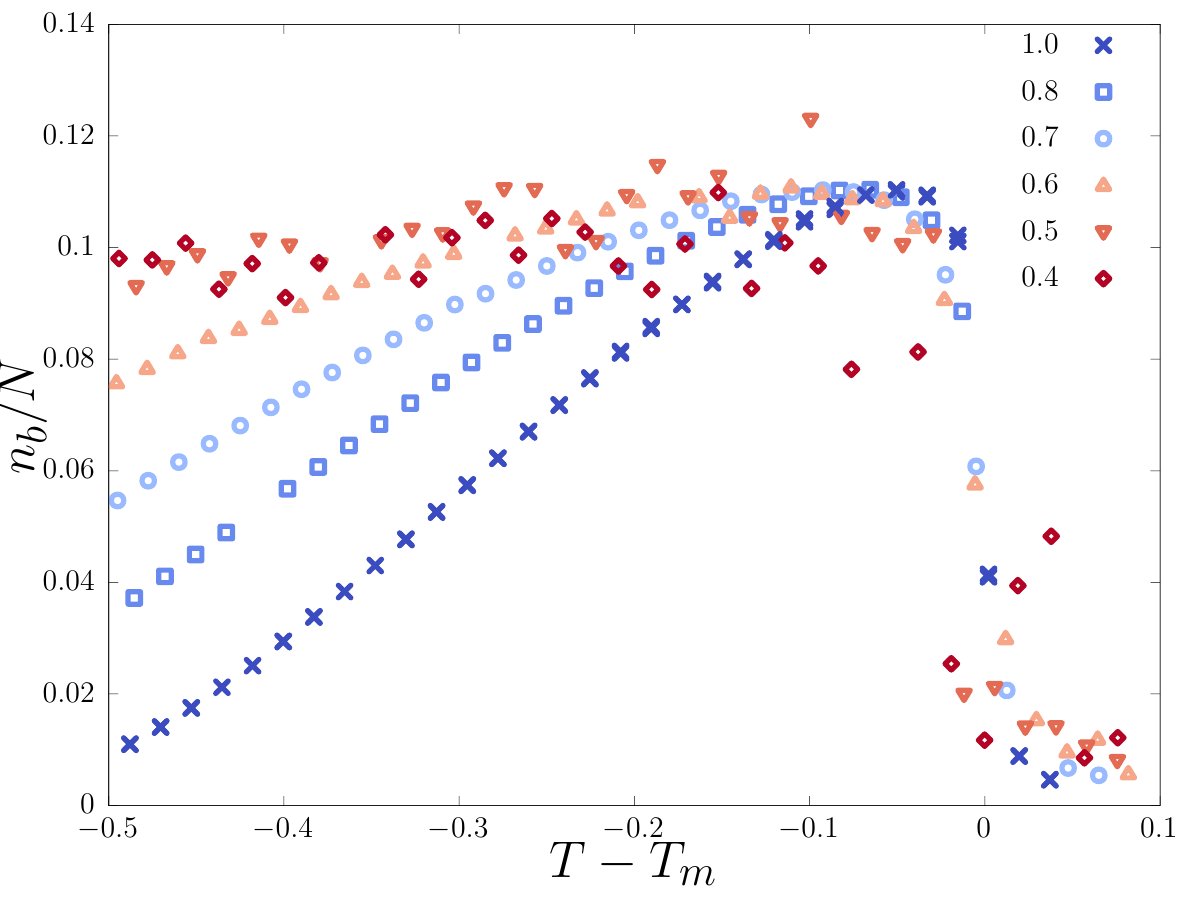}
\caption{\hl{Average number of bubbles per unit length $(n_b/N)$ across melting for different $p$ values. To compare the gradual reduction in $n_b$ away from $T_m$, we shift the x-axis by the corresponding melting points for each $p$.}}
\label{nbfig}
\end{figure}

\hl{Next, we look at the average number of bubbles. Bubble formation starts as the strands come close to each other, forming base pairs, and grows in number as $T$ is lowered below $T_m$. The number of bubbles $(n_b)$ peak close to $T_m$, and should gradually go to $n_b=0$ as $T\rightarrow 0$ in a tightly bound DNA \cite{majumdar2020}. For disordered lattices, we found $n_b$ to be higher for higher disorder when compared at equal distances from the corresponding $T_m$, indicating higher bubble stability concerning an equal temperature decrease from the melting point [Fig. \ref{nbfig}]. However, when compared at an absolute temperature $T$, $n_b$ is suppressed for systems with higher disorder. This change in the $n_b$ close to melting can have significant physical implications, e.g., in reality, bubbles are made of flexible single-stranded segments, therefore, how $n_b$ reduces below $T_m$ could significantly affect how the DNA looses rigidity while approaching the melting point \cite{majumdar2020}. Higher $n_b$ is expected to induce softness in the rigid bound state while also promoting bubble initiated processes.}

\begin{figure}[t]
\includegraphics[width=.8\linewidth]{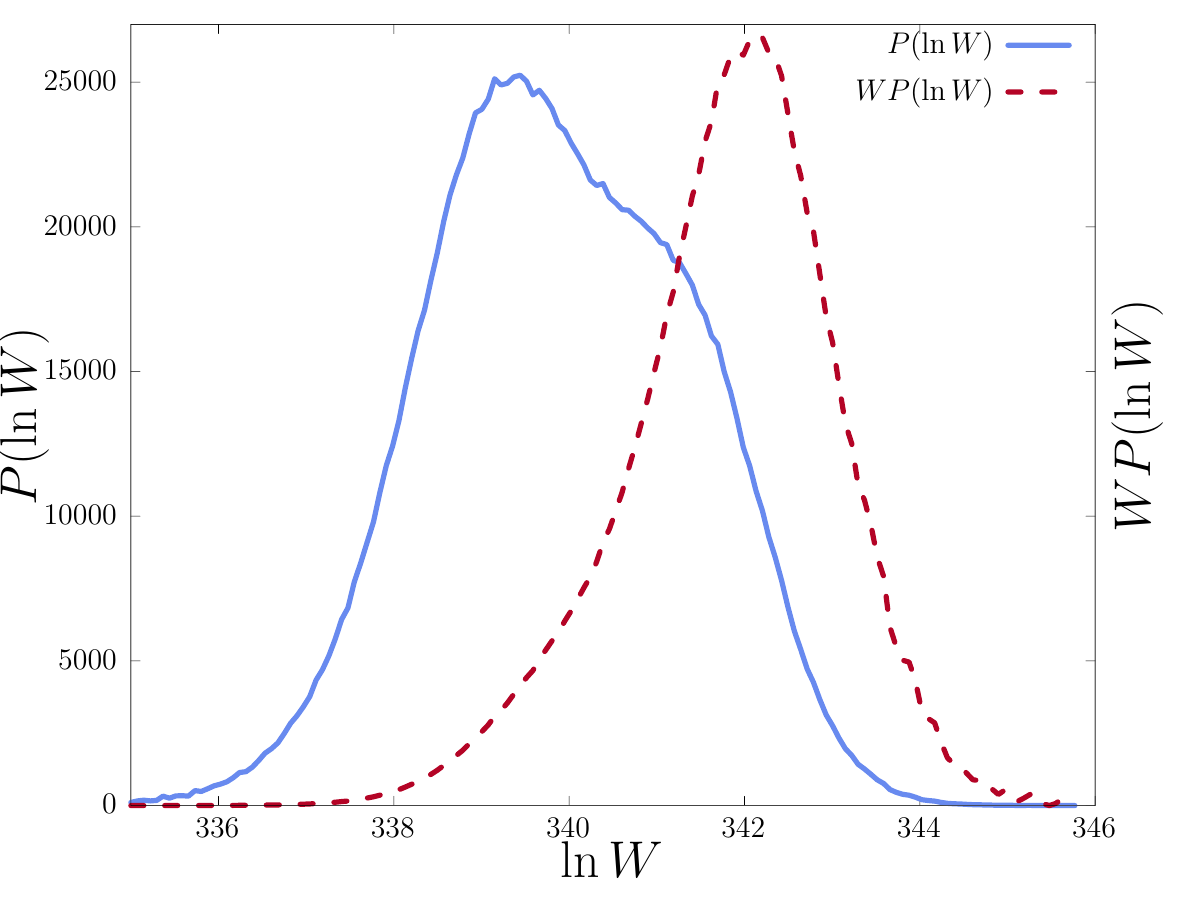}
\caption{Tour weight distribution $P(\ln W)$ and rescaled weighted distribution $WP(\ln W)$ at $T=0.895$ for $p=0.8$. The weight $W$ is exact up to a small multiplicative constant to contain numerical overflow.}
\label{plnwnfig}
\end{figure}

\textit{Validation and performance:} Finally, to validate our implementation of the PERM algorithm, we checked that our code reproduces results identical to Ref. \cite{causo2000} for the pure undiluted lattice $p=1$. Other than that, one of the major challenges in dealing with disordered backgrounds could be insufficient sampling, where few configurations contribute largely to the statistics. To show that our simulations are not glitched by such incomplete sampling, we show the contributions of each sample to the partition sum by comparing the distribution of tour weights $P(\ln W)$, and its weighted distribution $WP(\ln W)$ as suggested by Grassberger Ref. \cite{grassberger1999}, for $p=0.8$ at $T=0.895$ in Fig. \ref{plnwnfig}. Note that, $WP(\ln W)$'s contribution is large where $P(\ln W)$ is appreciable, ensuring the correctness of our results.

%CONCLUSION
\section{Conclusion \label{conclusion}}

In conclusion, we studied the effect of macromolecular crowders modeled as \hl{quenched disordered lattice sites}, on the melting of a lattice DNA model. Our findings demonstrate that crowders stabilize the double-stranded bound phase against thermal fluctuations \hl{leading to an increase in the melting temperature. The dependence of the melting temperature on disorder, however, has two parts: a nearly linear increase with disorder, followed by a strong non-linear increase.} Melting remains a first-order transition with no substantial change in the order parameter scaling exponent and the  bubble size exponent \hl{in the weak disorder regime, which, however, seems to change in the strong disorder regime. We plan to quantify this change in the scaling exponent in  our upcoming work.} Quenched and annealed type averaging showed no substantial difference in the order parameter. \hl{Also, the depth and breadth-first approach of disorder generation gives identical results.}

The most dramatic effect is perhaps the change in the probability distribution of the base-pair contacts near the melting point, which reveals that, for a sufficiently disordered environment, the states in between the bound and unbound phases are suppressed during the transition. \hl{This, however, is not something unexpected since the higher the disorder, the lesser the number of possible configurations of complex topology with bound phase and intermittent bubbles coexisting. Rather, having a topologically one-dimensional linear chain in either bound or unbound form is easier. Other than that, we also found the disorder to suppress the number of bubbles. However, bubbles become resilient when compared at equal distances from the corresponding melting points for different disorder values.} 

\hl{Our results, in a way, corroborate the importance of excluded volume interaction for DNA in particular and biophysical processes in general. While \textit{in vitro}, controlled dissociation of DNA strands is carried out simply by varying the temperature or \textit{p}H of the DNA solution, such maneuvers are infeasible physiologically, which makes the crowded environment to stand out as a potential candidate to alter the stability of the DNA duplex structure, thereby making it biologically important.}

Lastly, while disorder comes in many different forms, we consider the simplest possible case, which includes \hl{chemically inert, interacting only by volume exclusion, and spatially }non-correlated type, for our study. Depending upon the electrostatic interaction with the crowders, the disorder can be attractive too, with long-range correlations \cite{haydukivska2014}. Such considerations are currently underway, which we aim to publish them in the future. 

%ACKNOWLEDGEMENT
\section{Acknowledgement}
DM would like to thank Deepak Dhar, Somendra M Bhattacharjee, Peter Grassberger, Soheli Mukherjee, and Dibyajyoti Mohanta for insightful discussions and valuable feedback during the research work. DM was supported by the BCSC Fellowship from the Jacob Blaustein Center for Scientific Cooperation and by the Israel Science Foundation (ISF) through Grant No. 1301/17 and 1204/23. 

%BIBLIOGRAPHY
%\bibliographystyle{unsrt}
%\bibliography{fractal_melting.bib}

\begin{thebibliography}{99}

% 1 - 10
\bibitem{fulton1982} Fulton, A. B. How crowded is the cytoplasm? \textit{Cell} \textbf{1982},  30, 345–347.

\bibitem{miyoshi2008} Miyoshi, D.; Sugimoto, N. Molecular crowding effects on structure and stability of DNA. \textit{Biochimie} \textbf{2008}, 90, 1040–1051.

\bibitem{skora2020}  Sk\'ora, T.; Vaghefikia, F.; Fitter, J.; Kondrat, S. Macromolecular crowding: How shape and interactions affect diffusion. \textit{J. of Phys. Chem. B} \textbf{2020}, 124, 7537–7543.

\bibitem{singh2022} Singh, A.; Maity, A.; Singh, N. Structure and dynamics of dsDNA in cell-like environments. \textit{Entropy} \textbf{2022}, 24, 1587.

\bibitem{neha2024} Mathur, N.; Singh, A.; Singh, N. Force-induced unzipping of DNA in the presence of solvent molecules. \textit{Biophysical Chemistry} \textbf{2024}, 307, 107175.

\bibitem{comm1} Atmosphere refers to the immediate environment of the DNA.

\bibitem{liu2010} Liu, Y.; Kermanpour, F.; Liu, H. L.; Hu, Y.; Shang, Y. Z.; Sandler, S. I.; Jiang, J. W. Molecular thermodynamic model for DNA melting in ionic and crowded solutions. \textit{J. Phys. Chem. B} \textbf{2010}, 114, 9905-9911.

\bibitem{singh2017} Singh, A.; Singh, N. DNA melting in the presence of molecular crowders. \textit{Phys. Chem. Chem. Phys.} \textbf{2017} 19, 19452.

\bibitem{hong2020}  Hong, F.;  Schreck, J. S.; \v{S}ulc, P. Understanding DNA interactions in crowded environments with a coarse-grained model. \textit{Nucl. Acds. Res.} \textbf{2020}, 48, 19.

\bibitem{woolley1985}  Woolley, P.; Wills, P. R. Excluded-volume effect of inert nucleic acids macromolecules on the melting of nucleic acids. \textit{Biophy. Chem.} \textbf{1985}, 22, 89–94.

% 11 - 20
\bibitem{nakano2004} Nakano, S.; Karimata, H.; Ohmichi, T.; Kawakami, J.; Sugimoto, N. The effect of molecular crowding with nucleotide length and cosolute structure on DNA duplex stability. \textit{J. Am. Chem. Soc.} \textbf{2004}, 126, 14330-14331.

\bibitem{harve2009} Harve, K. S.; Lareu, R.; Rajagopalan, R.; Raghunath, M. Understanding how the crowded interior of cells stabilizes DNA/DNA and DNA/RNA hybrids - \textit{in silico} predictions and \textit{in vitro} evidence. \textit{Nucl. Acds. Res.} \textbf{2010} 38(1), 172-181.

\bibitem{wieder1981} Wieder, R.;  Wetmur, J. G. One hundred-fold acceleration of DNA  renaturation rates in solution. \textit{Biopolymers} \textbf{1981}, 20, 1537–1547.

\bibitem{sikorav1991} Sikorav,  J.-L.; Church, G. M. Complementary recognition in condensed DNA: Accelerated DNA renaturation. \textit{J. Mol. Biol.} \textbf{1991}, 222, 1085–1108.

\bibitem{goobes2003}  Goobes, R.;  Kahana, N.; Cohen, O.;  Minsky, A. Metabolic buffering exerted by macromolecular crowding on DNA-DNA interactions: Origin and physiological significance. \textit{Biochemistry} \textbf{2003}, 42, 2431–2440.

\bibitem{pbd1989} Peyrard, M.; Bishop, A. R. Statistical mechanics of a nonlinear model for DNA denaturation. \textit{Phys. Rev. Letts.} \textbf{1989}, 62, 2755.

\bibitem{bublee1988}  Lee, S. B.; Nakanishi, H. Self-avoiding walks on randomly diluted lattices. \textbf{Phys. Rev. Letts.} \textbf{1988}, 61, 18.

\bibitem{meir1989} Meir, Y.;  Harris, A. B. Self-avoiding walks on diluted networks. \textit{Phys. Rev. Letts.} \textbf{1989}, 63, 26.

\bibitem{rintoul1994} Rintoul, M. D.;  Moon, J.; Nakanishi, H. Statistics of self-avoiding walks on randomly diluted lattices. \textit{Phys. Rev. E} \textbf{1994}, 49, 4.

\bibitem{singh2009} Singh, A. R.;  Giri, D.; Kumar, S. Effects of molecular crowding on stretching of polymers in poor solvent. \textit{Phys. Rev. E} \textbf{2009} 79, 051801. 

% 21 - 30
\bibitem{blavatska2010}  Blavatska, V.; Janke, W. Shape anisotropy of polymers in disordered environment. \textit{J. Chem. Phys.} \textbf{2010}, 133, 184903.

\bibitem{stauffer1992} Stauffer, D. Introduction to percolation theory. Taylor \& Francis, London, 2nd edition.

\bibitem{metze2013}  Metze, K. Fractal dimension of chromatin: potential molecular diagnostic applications for cancer prognosis. \textit{Expert Review of Molecular Diagnostics} \textbf{2013}, 13(7), 719–735.

\bibitem{tamm2015} Tamm, M. V.; Nazarov, L. I.; Gavrilov, A.  A.; Chertovich, A. V. Anomalous diffusion in fractal globules. \textit{Phys. Rev. Lett.} \textbf{2015}, 114, 178102.

\bibitem{weber2011}  Weber, S. C.; Spakowitz, A. J.; Theriot, J. A. Nonthermal ATP-dependent fluctuations contribute to the in vivo motion of chromosomal loci. \textit{PNAS} \textbf{2012}, 109(19), 7338–7343.

\bibitem{singh2024} Singh, S.; Granek, R. Active fractal networks with stochastic force monopoles and force dipoles unravel sub-diffusion of chromosomal loci. \textit{Chaos} \textbf{2024}, 34, 113107.

\bibitem{majumdar2024} Majumdar, D.; Singh, S.; Granek, R. (unpublished).

\bibitem{causo2000} Causo, M. S.; Coluzzi, B.; Grassberger, P. Simple model for the DNA denaturation transition. \textit{Phys. Rev. E} \textbf{2000}, 62(3), 3958-3973.

\bibitem{poland1966} Poland, D.;  Scheraga, H. A. Phase transitions in one dimension and the helix-coil transition in polyamino acids. \textit{J. Chem. Phys.} \textbf{1966}, 45, 1456–1463.

\bibitem{geggier2010} Geggier, S.; Vologodskii, A. Sequence dependence of DNA bending rigidity. \textit{Proc. Natl. Acad. Sci.} \textbf{2010}, 107, 15421.

\bibitem{yuan2006} Yuan, C.; Rhoades, E.; Lou, X. W.; Archer, L. A. Spontaneous sharp bending of dna: Role of melting bubbles. \textit{Nucl. Acds. Res.} \textbf{2006}, 34, 4554.

\bibitem{titus2005} Erp, van T. S.; Cuesta-Lopez, S.; Hagmann, J-G; Peyrard, M. Can one predict DNA transcription start sites by studying bubbles? \textit{Phys. Rev. Letts.} \textbf{2005}, 95, 218104.

\bibitem{majumdar2020} Majumdar, D.; Bhattacharjee, S. M. Softening of DNA near melting as disappearance of an emergent property. \textit{Phys. Rev. E} \textbf{2020}, 102, 032407.

\bibitem{majumdar2021} Majumdar, D. Elasticity of a DNA chain dotted with bubbles under force. \textit{Phys. Rev. E} \textbf{2021}, 103, 052412.

\bibitem{majumdar2023} Majumdar, D. Adsorption of melting deoxyribonucleic acid. \textit{Phys. of Fluids} \textbf{2023}, 35, 067110.

\bibitem{majumdar2023p2} Majumdar, D. DNA melting in poor solvent. \textit{J. Stat. Phys.} \textbf{2023}, 190, 14.

\bibitem{coluzzi2006} Coluzzi, B. Numerical study of a disordered model for DNA denaturation transition. \textit{Phys. Rev. E} \textbf{2006}, 73, 011911.

\bibitem{leath1976} Leath, P. L. Cluster size and boundary distribution near percolation threshold. \textit{Phys. Rev. B} \textbf{1976}, 14, 5046.

\bibitem{hsu2005}  Hsu, H-P.; Nadler, W.; Grassberger, P. Simulations of lattice animals and trees. \textit{J. Phys. A: Math. Gen.} \textbf{2005}, 38, 775–806.

\bibitem{grassberger1997} Grassberger, P. Pruned-enriched Rosenbluth method: Simulations of $\theta$ polymers of chain length up to 1 000 000. \textit{Phys. Rev. E} \textbf{1997}, 56(3), 3682-3693.

\bibitem{bachmann2004} Bachmann, M.; Janke, W. Thermodynamics of lattice heteropolymers. \textit{J. Chem. Phys.} \textbf{2004}, 120, 14.

\bibitem{rosenbluth1955} Rosenbluth, M. N.; Rosenbluth, A. W. Monte Carlo calculation of the average extension of molecular chains. \textit{J. Chem. Phys.} \textbf{1955}, 23, 356.

\bibitem{matsumoto1998} Matsumoto, M.; Nishimura, T. Mersenne twister: A 623-dimensionally equidistributed uniform pseudo-random number generator. \textit{ACM Transactions on Modeling and Computer Simulation} \textbf{1998}, 8(1), 3–30.

\bibitem{numericalrecipes} Press, W. H.; Teukolsky, S. A.; Vetterling, W. T.; Flannery, B. P. Numerical Recipes in C: The art of Scientific Computing. Cambridge University Press, 2002.

\bibitem{frauenkron1999} Frauenkron, H.; Causo, M. S.; Grassberger, P. Two-dimensional self-avoiding walks on a cylinder. \textit{Phys. Rev. E} \textbf{1999}, 59(1).

\bibitem{carlon2002} Carlon, E.; Orlandini, E.; Stella, A. L. Roles of stiffness and excluded volume in DNA denaturation. \textit{Phys. Rev. Letts.} \textbf{2002}, 88(19).

\bibitem{doussal1991} Doussal, P. L.; Machta, J. Self-avoiding walks in quenched random environments. \textit{J. Stat. Phys.} \textbf{1991}, 64(3/4).

\bibitem{cherayil1990} Cherayil, B. J. Equilibrium dimensions of polymers in quenched disorder. \textit{J. Chem. Phys.} \textbf{1990}, 92, 6246.

\bibitem{wu1991} Wu, D.; Hui, K.; Chandler, D. Monte Carlo study of polymers in equilibrium with random obstacles. \textit{J. Chem. Phys.} \textbf{1991} , 96, 835.

\bibitem{blavatska2013}  Blavatska, V. Equivalence of quenched and annealed averaging in models of disordered polymers. \textit{J. of Phys. Cond. Matt.} \textbf{2013}, 25, 505101.

\bibitem{bradly2021} Bradly, C. J.; Owczarek, A. L. Effect of lattice inhomogeneity on collapsed phases of semi-stiff ISAW polymers. \textit{J. Stat. Phys.} \textbf{2021}, 182, 27.

\bibitem{haydukivska2016} Haydukivska, K.; Blavatska, V. Loop statistics in polymers in crowded environment. \textit{J. Chem. Phys.} \textbf{2016}, 144, 084901.

\bibitem{grassberger1999} Grassberger, P. Comment on “polymer localization in attractive random media”. \textit{J. Chem. Phys.} \textbf{1999}, 111, 440–442.

\bibitem{haydukivska2014} Haydukivska, K.; Blavatska, V. Ring polymers in crowded environment: Conformational properties. \textit{J. Chem. Phys.} \textbf{2014}, 141, 094906.

\end{thebibliography}
%\end{document}

\end{document}